\documentclass{article}

\usepackage{PRIMEarxiv}

\usepackage[utf8]{inputenc} 
\usepackage[T1]{fontenc}    
\usepackage{hyperref}       
\usepackage{url}            
\usepackage{booktabs}       
\usepackage{amsfonts}       
\usepackage{nicefrac}       
\usepackage{microtype}      
\usepackage{lipsum}
\usepackage{fancyhdr}       
\usepackage{graphicx}       
\graphicspath{{media/}}     
\usepackage{listings}
\usepackage{xcolor}
\usepackage{wrapfig}

\title{Dynamic Code Orchestration: Harnessing the Power of Large Language Models for Adaptive Script Execution
}

\author{
  Justin Del Vecchio, Andrew Perrault, Eliana Furmanek \\
  Canisius University\\
  Buffalo, NY\\
  \texttt{\{delveccj, perreaua, furmanee\}email@canisius.edu} \\
}

\begin{document}
\maketitle

\begin{abstract}
Computer programming initially required humans to directly translate their goals into  machine code.  
These goals could have easily been expressed as a written (or human) language directive.  Computers, however, had no 
capacity to satisfactorily interpret written language. Large language model's provide exactly this capability; 
automatic generation of computer programs or even assembly code from written language directives.  
This research examines dynamic code execution of written language directives within the context
of a running application.  It implements a text editor whose business logic is purely backed by 
large language model prompts.  That is, the program's execution uses prompts and written language 
directives to dynamically generate application logic at the point in time it is needed.  
The research clearly shows how written language directives, backed by a large language model, offer 
radically new programming and operating system paradigms.  For example, empowerment of users to directly 
implement requirements via written language directives, thus supplanting the need for a team of
programmers, a release schedule and the like. Or, new security mechanisms where static executables, 
always a target for reverse engineering or fuzzing, no longer exist.  They are replaced by ephemeral executables that may continually change, be completely removed, and are easily updated. 
\end{abstract}


\section{Introduction}
\begin{figure}[h]
    \centering
    \includegraphics[width=1\columnwidth]{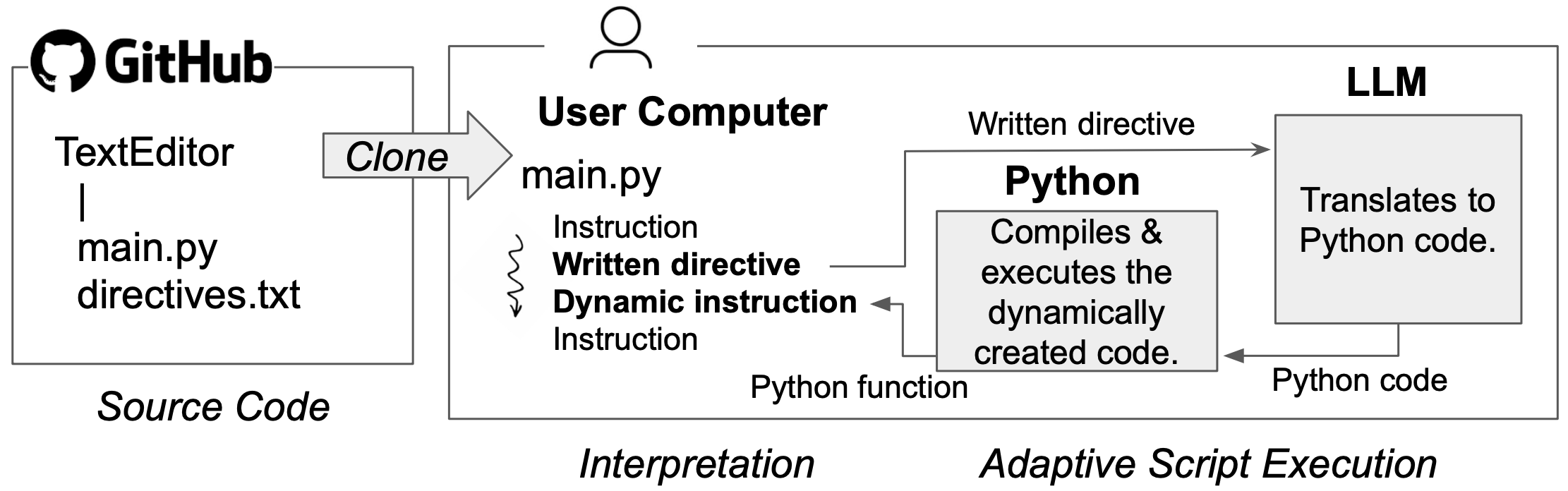}  
    \caption{Dynamic code orchestration data flow.}
    \label{fig:figure1}
\end{figure}

Imagine a future where software projects are distributed and used as shown in Figure~\ref{fig:figure1}. A description of the Figure's key elements follows.

\textit{Source Code} - Project source code exists in a high level language.  However, it serves as a skeleton
    that enables application loading and limited functionality.  Much of the application's business logic 
    exists as a set of written language directives.  An example of such a directive is provided in 
    Section~\ref{sec:DCO}.
    
\textit{Interpretation or Compilation} - Users configure the project to use a large
    language model (LLM) service.  This may be installed locally or available via a network.  
    The project skeleton code is interpreted or compiled with hooks to access the LLM 
    when required by the application.

\textit {Adaptive Script Execution} - A user runs the application and interacts with it.  Written
    language directives embody large portions of the application's business logic.  When this logic
    is required, the directives are sent to the LLM service and returned as Python code.  The code is 
    compiled into an executable code block and registered as function in the global namespace .  
    This function may then be called, or completely regenereated, as often as needed.

Dynamic code orchestration transforms applications into a custom patchwork of ephemeral executable code 
blocks represented by adaptive written directives.  The advantages of this approach include:

\begin{itemize}
    \item \textbf{User Customization}: User's can define custom functionality as a written directive.
    Applications can be extended by user's themselves.  This removes the need for a team of supporting
    programmers who interpret requirements and develop code.  
    \item \textbf{Reduced Attack Surface}: Fuzzing and reverse engineering an executable becomes 
    harder as business logic is encapsulated in written language directives.  LLMs can be instructed
    to return syntactically different but semantically equivalent executable code blocks.  This makes
    it difficult for malware developers to target vulnerabilities as applications are no longer
    statically defined.
    \item \textbf{Space Conservation}: Code executables become far less bulky.  Unused code simply does not exists.  The written language directives dictate what composes an executable.  The executable
    code blocks generated by a written directive can periodically be purged to conserve space.
\end{itemize}

We examined the dynamic code orchestration concept via two research questions.  We provide the code developed to answer both questions as public GitHub repositories.

1. Could we develop an application that is dynamically orchestrated?  We 
would need to bake into the application a capability to call an LLM service.  Certain user
actions would lead to written language directives.  The directives 
would need to be sent to the LLM, executable code blocks generated and added to the global
namespace.  The functions could then be called by the application.  This 
question would require use of prompt engineering \cite{LangChain} and 
custom LLM knowledge stores that used our application source code for context when formulating replies.  

2. What is the current veracity of written language directive to code translation?  How well
do LLMs currently perform at this task?  We used the HumanEval set \cite{Chen2021} as 
a benchmark.  This benchmark has been tested against ChatGPT 3.5 and 4 models, amongst
others.  Our goal was to identify what failures (incorrect or uncompilable LLM replies) 
looked like.  How would our dynamic code orchestration pipleline 
need to adjust for such failures?  Additionally, how could the executable code blocks be
seamlessly integrated into a running application, in this case the test harness application?

\section{Q1: Implement a Dynamic Code Orchestration Application}\label{sec:DCO}

\begin{figure}[h]
    \centering
    \includegraphics[width=1\columnwidth]{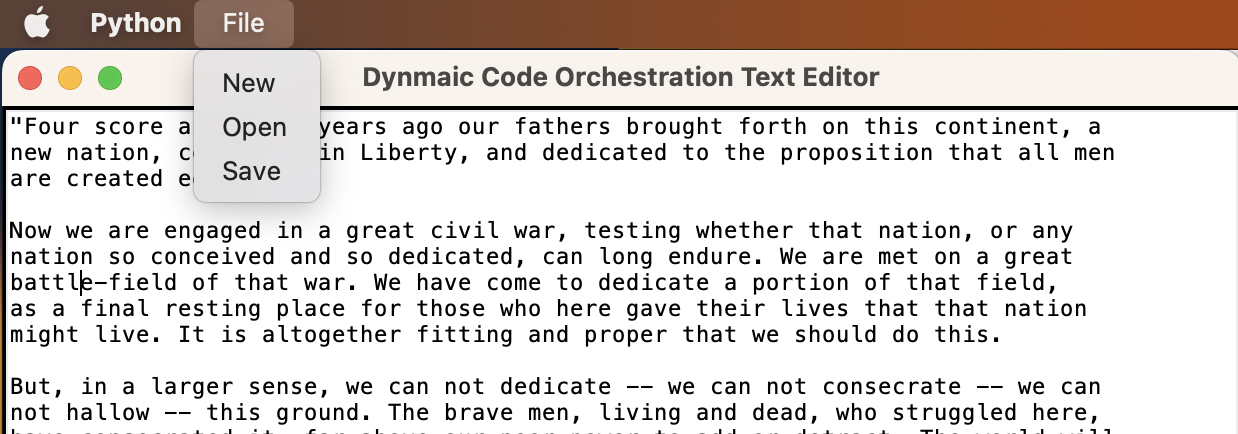}  
    \caption{Dynamic code orchestration text editor.}
    \label{fig:figure2}
\end{figure}

We created a dynamically orchestrated Python text editor.  The application allows
users to open, edit, and save text documents.  The user interface has a drop down menu 
with items for \textbf{New}, \textbf{Open}, and \textbf{Save}.  \textbf{New} creates an entirely 
new file, \textbf{Open} opens an existing file, and \textbf{Save} saves the file currently 
open in the editor.  The editor itself is shown in Figure~\ref{fig:figure2}.

\begin{figure}[h]
    \centering
    \includegraphics[width=1\columnwidth]{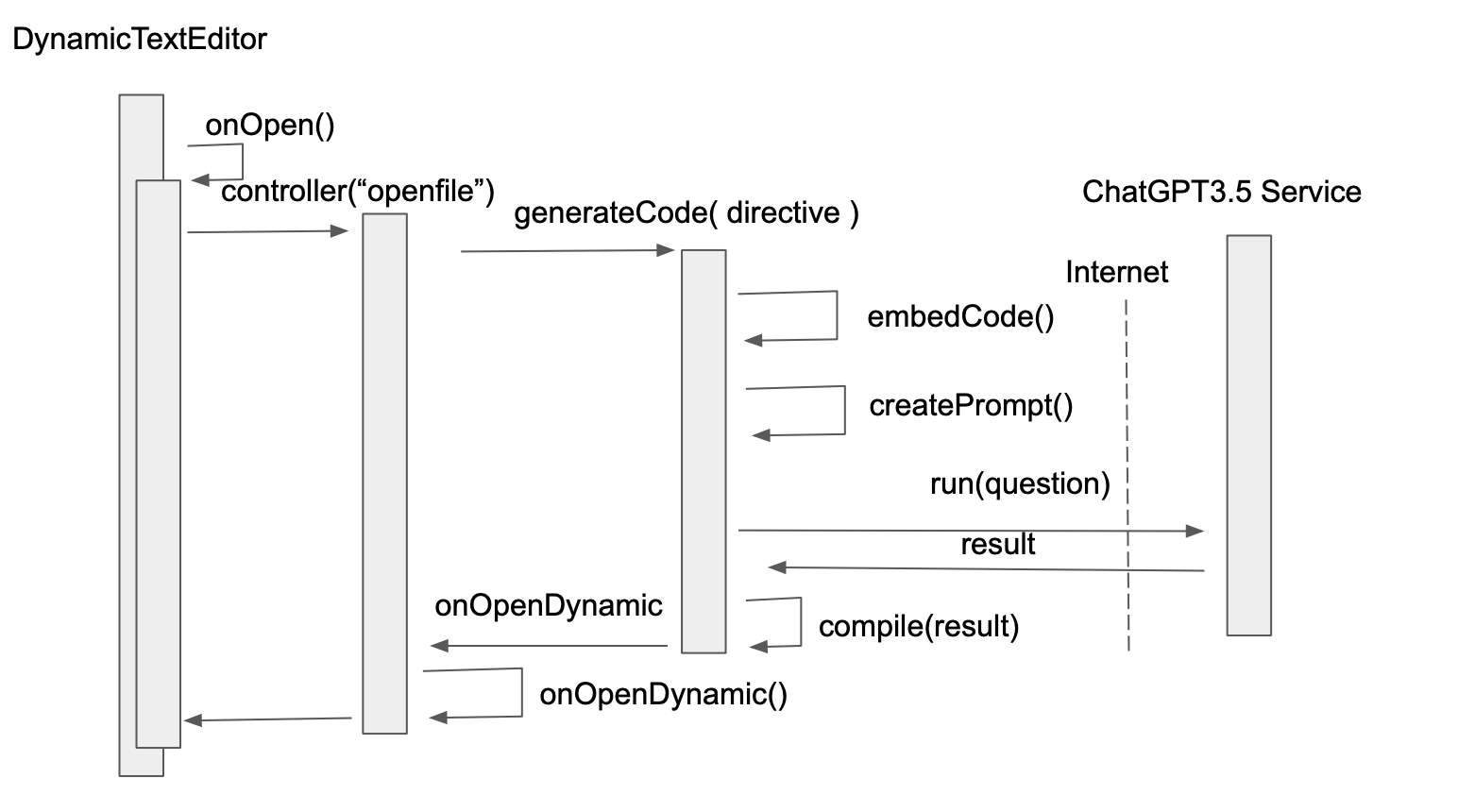}  
    \caption{Dynamic code orchestration sequence diagram.}
    \label{fig:figure3}
\end{figure}

The skeleton code for the application created the GUI elements for menu items.  It added 
the elements to a layout and established a command for each menu item.  When a user clicked on 
a menu item, the dynamic code orchestration workflow would initiate.  The sequence of events is 
shown in Figure~\ref{fig:figure3} with a description following. Here, the user has clicked \textbf{Open}. 

Dynamic Code Orchestration Event Sequence

\textbf{Step 1}. The \texttt{controller} function is called.  It is passed a command that it is
to prepare to open a file.  It finds the requisite written directive for the 
desired executable code block.  The contents of the directive are discussed in \textbf{Step 4}.
It then calls \texttt{generateCode} and passes the directive.

\textbf{Step 2}. \texttt{generateCode} checks to see if this is the first time dynamic code has
been requested.  If so, it embeds the existing source code of \texttt{DynamicTextEditor} 
in terms of the ChatGPT 3.5 model.  Think of the embed step as creating a customized
\cite{OpenaAIFineTuning} question and answer chat model that is aware of the
content of \texttt{DynamicTextEditor} script.  Further, the custom model recognizes that 
this file is Python source code.  All questions to the model will be made in the context 
of this source code.

\textbf{Step 3}. \texttt{generateCode} next creates a prompt which can pose questions to the custom
ChatGPT 3.5 model.  Following is the text of the prompt used in the application:

\texttt{\textbf{LLM Prompt:}} \textit{"You are a programmer. You should use the preexisting code in 
the file DynamicTextEditor.py and create the requested functions so the code operates without error.
Pay attention to the imports in DynamicTextEditor and choose code that works within those imports."}

\textbf{Step 4}. Next, the written directive for \texttt{openFile} is sent to the LLM's 
question and answering mechanism.  Following is the text of the written directive that
instructs on how to open a file:

\texttt{\textbf{LLM Open File Directive:}} \textit{"Create a single function named onOpenDynamic.  It takes one argument, self. The function uses a file dialog to let the user select and open a file. The file is then placed in the Text object in the initUI() function in for editing. Include inside the function any necessary imports to make the code run."}

\textbf{Step 5}. The question is sent to the ChatGPT 3.5 model over the 
Internet.  Note, this model could just as easily be locally called using alternative
LLM solutions such as Meta's Llama\cite{Llama}.  

\textbf{Step 6}. The response is returned.  The following listing shows the reply for the
open file directive.   

\begin{lstlisting}[basicstyle=\small]
def onOpenDynamic(self):
    from tkinter import filedialog
    filename = filedialog.askopenfilename(initialdir = "/",title = "Select file",filetypes = (("Text files","*.txt"),("all files","*.*")))
    self.text.delete('1.0', END)
    with open(filename, 'r') as f:
        self.text.insert(END, f.read())\end{lstlisting}

The code response does exactly what is needed based on the written directive.  It creates a dialog
with which the user may open the file and replaces the current content of the text area with the 
selected file.  Note, LLMs allow us to adjust the model's drift so that a wide range of 
semantically equivalent but syntactically different source code replies may be provided. As well, we could request the exact same source code reply be provided 
each time for a particular question.

\textbf{Step 7}. The returned code is compiled with \texttt{compile} into a code object
that can be executed by the Python interpreter. It does not produce standalone executables
or machine code like a traditional compiler; rather, it generates a code object that can be
executed within the current Python session.  We store the code object as a callable function
in the application's global namespace.  Our technique to do this was perfected in the research question 
detailed in Section~\ref{sec:Fails} and is described there.

\textbf{Step 8}.  The callable function is returned to the \texttt{controller} function
and invoked.  A file dialog is presented to open a file.  If a file is selected, it is 
then placed in the application's text area. 

\subsection{User Defined Functionality}
There is a potential problem with \textbf{Step 6}.  What if text is already present in the 
text area?  The returned function would simply overwrite the text.  The user might want to
save the current file before opening a new one.  It would be nice to warn the user and allow them to 
cancel the operation!

We edited the directive for \textbf{openFile} while the application ran. The directive was 
changed with the following text added:

\texttt{\textbf{LLM Open File Directive Enhancement:}} \textit{"If there is already content in the text area - warn the user!"}

The next menu click of \textbf{Open} generated the following code:
\begin{lstlisting}[basicstyle=\small]
def onOpenDynamic(self):
    import tkinter as tk
    from tkinter import filedialog
    from tkinter import messagebox
    if self.text.get("1.0", "end-1c") != "":
        messagebox.showwarning("Warning", "There is already content in the text area. Opening a new file will overwrite the existing content.")
    file_path = filedialog.askopenfilename()
    if file_path != "":
        with open(file_path, "r") as f:
            self.text.delete("1.0", tk.END)
            self.text.insert("1.0", f.read())
            \end{lstlisting}

\begin{figure}
  \centering
  \includegraphics[width=0.2\textwidth]{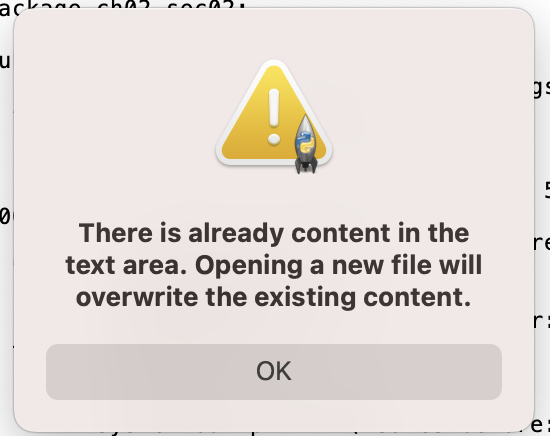} 
  \caption{Dynamic warning.}
 \label{fig:figure4}
\end{figure}

ChatGPT augmented the code to add a message box warning of the potential issue.  The user now
has the option to cancel the operation.  This new functionally required no editing of the
original source code yet integrated with it seamlessly.  Figure~\ref{fig:figure4}
shows what we saw after a simple update to the written directive.

\section{Q2: Failure Points of Generated Code}\label{sec:Fails}

We performed an experiment to see how well the ChatGPT LLM generated function code
for an arbitrary written directive.  We utilized the HumanDataEval
dataset\cite{Chen2021}.  It contains 164 algorithm based problems, as well as unit tests,
and a proper answer to the problem. We judged success if the generated function would pass a set number of
unit tests. 

We wrote code which prompted ChatGPT with a written directive task via its API.  We generated
10 different versions of a single function to satisfy each of the 164 problems. ChatGPT was provided a
function header, the problem description, and a format for its JSON result.  We needed to provide
the result format to clearly distinguish the developed code from uncommented descriptions of code 
functionality ChatGPT might add.

Our test harness wrote the ChatGPT replies to file, imported the code
dynamically, and tested it against the unit tests. We implemented a technique that dynamically
translated ChatGPT replies into executable code blocks.  The approach was used by the text 
editor detailed in Section~\ref{sec:DCO}.  The following code listing provides a condensed view of the technique.

\begin{lstlisting}[basicstyle=\small]
1 result = dcc_qa.run(question[0])
2 compiled_code = compile(result, filename="", mode="exec")
3 locals_dict = {}
4 exec(compiled_code, globals(), locals_dict)
5 test_function = locals_dict.get("test24", None)
6 if test_function is None or not callable(test_function):
7    print(f"Function not found or not callable.")
8    return
9 test_function(self)
\end{lstlisting}

Line 1 submits the written narrative to the ChatGPT LLM as a question.  
It replies with the source code for the Python function.  Line 2 compiles 
the code into a code object. Line 4 is the critical step.  
The \texttt{exec} function is called on the \texttt{compiled\_code} code object. 
Specifically, any functions 
defined within the \texttt{compiled\_code} will be registered in the global namespace 
(\texttt{globals()}). This means that these functions will become globally accessible 
after the \texttt{exec()} call. Here's a breakdown of how it works\cite{ChatGPTGlobals}:

\begin{enumerate}
  \item The \texttt{exec()} function is used to execute the Python code contained in \texttt{compiled\_code}.
  \item When this code is executed, any functions defined within it will be registered in the 
  global symbol table (namespace) because \texttt{globals()} is used as the second argument.
  \item Once registered in the global namespace, these functions can be accessed and called from anywhere within the program.
\end{enumerate}

Lines 5 through 8 test if the code block generated a legitimate, executable function.  Line 9 
invokes the newly created function which is now available in the global namespace.

We  tracked and logged the output and results of generating code block executables for the 
HumanEval test set.  Using 1400 test files, ChatGPT 3.5 successfully generated code that compiled and ran 
correctly 61\% of the time. This number would likely greatly improve with a switch
to the ChatGPT 4 model which we plan to use in the future.

\subsection{Challenges for Dynamically Generate Code }
One challenge is getting ChatGPT to respond with a consistent JSON format
where it is easy to identify what is source code and what is not. ChatGPT was instructed to explicitly delimit
code with three back ticks.  However, it did not always correctly do this. 15\% of the time we were 
unable to confidently isolate the  code in the response (thus the reason we ran 1400 tests as opposed to 
1640).  This problem could be overcome through imporved identification and removal of uncommented code
descriptions added by ChatGPT. 

Another challenge was getting ChatGPT to not import from external libraries as we
could not guarantee they would be available. Even when instructed not to import 
external libraries it still would. This problem could be overcome through a fine tuned LLM 
model or by having \texttt{!pip install newlibrary} commands added for any imported libraries.

A final challenge we faced was the LLM writing infinite while loops, clearly the greatest
problem. Since the HumanEval problems are relatively simple, we implemented a timeout if the
function ran pass a time threshold.  Part of our future research will include 
experiments where we specifically instruct the LLM to protect against infinite loops and measure its 
performance with and without such guidance.

\section{Related Work}

Research into pseudo code languages that allow specification of code for real time compilation exist \cite{Poletto1999}\cite{Engler}.   As well, Just in Time Compilers (JIT) have used the technique. These approaches have not considered an LLM that can synthesize source code or assembly from a modest set of natural language directives.  Natural language instructions to generate source code have also been studied.  A survey of approaches from 2021 is found here \cite{Shin2021}.  These 
mainly focus on controlled vocabulary that map to source code constructs.

Researchers performed a set of experiments on the now deprecated OpenAI platform “CodeX” \cite{Tony2023}. CodeX was a GPT-3 LLM that was designed specifically for programmers. It was designed to take in prompts, some code, and create programs. It was deprecated upon the release of ChatGPT as ChatGPT was deemed a much better resource.

\section{Future Plans}
We will focus on two research areas.  Specifically:

\textbf{User Defined Adaptive Scripts} - This research will focus on identifying the 
possibilities and limitations of user extended application functionality.  We will introduce 
written directives backed by cascading LLM prompts to extend an existing open source 
application or library.  We will review user requests for enhancements and select a set 
to implement with dynamic code orchestration, playing the part of empowered user.

\textbf{Defensive Coding} - We will take an open source project with a Windows 
distributable.  It will have past vulnerabilities identified as MITRE CVEs\cite{MITRE}.  
We will translate portions of the existing code base into written directives.  
We will then perform a security analysis of the application to identify vulnerabilities.  This will include fuzzing, disassembly, and debugging.

\bibliographystyle{unsrt}  
\bibliography{references}

\begin{thebibliography}{10}

\bibitem{LangChain}
LangChain.
\newblock Prompt engineering, 2023.
\newblock Accessed: Septemeber 9, 2023.

\bibitem{Chen2021}
Evaluating large language models trained on code.
\newblock 7 2021.

\bibitem{OpenaAIFineTuning}
OpenAI.
\newblock Fine tuning, 2023.
\newblock Accessed: Septemeber 8, 2023.

\bibitem{Llama}
Meta.
\newblock Meta llama, 2023.
\newblock Accessed: Septemeber 8, 2023.

\bibitem{ChatGPTGlobals}
OpenAI.
\newblock Globals conversation, 2023.
\newblock Accessed: Septemeber 8, 2023.

\bibitem{Poletto1999}
Massimiliano Poletto, Wilson~C Hsieh, Dawson~R Engler, M~Frans Kaashoek, :~M Poletto, D~R Engler, M~F Kaashoek, and ;~W~C Hsieh.
\newblock 'c and tcc: A language and compiler for dynamic code generation 'c and tcc: A language and compiler for dynamic code generation · 325, 1999.

\bibitem{Engler}
Dawson~R Engler, Wilson~C Hsieh, and M~Frans Kaashoek.
\newblock 'c: A language for high-level, efficient, and machine-independent dynamic code generation.

\bibitem{Shin2021}
Jiho Shin and Jaechang Nam.
\newblock A survey of automatic code generation from natural language.
\newblock {\em Journal of Information Processing Systems}, 17:537--555, 6 2021.

\bibitem{Tony2023}
Catherine Tony, Markus Mutas, Nicolás E.~Díaz Ferreyra, and Riccardo Scandariato.
\newblock Llmseceval: A dataset of natural language prompts for security evaluations.
\newblock 3 2023.

\bibitem{MITRE}
MITRE.
\newblock Mitre cve, 2023.
\newblock Accessed: Septemeber 8, 2023.

\end{thebibliography}

\end{document}